\documentclass[aps,prd
    ,nofootinbib
    ,groupedaddress
    ,twocolumn
    ,showpacs]{revtex4}

\usepackage{amsmath}
\usepackage{amsfonts}

\newcommand{\Or}{{O}}
\newcommand{\gO}{g^0}
\newcommand{\gm}[1]{\gamma_{#1}}
\newcommand{\calR}{{\cal R}}

\newcommand{\Deqn}[1]{{Eq.\ (\ref{#1})}}
\newcommand{\Deqns}[1]{{Eqs.\ (\ref{#1})}}

\newcommand{\rd}{{\mbox{d}}}
\newcommand{\dx}{{\mbox{d}x}}
\newcommand{\dt}{{\mbox{d}t}}
\newcommand{\s}{{\mbox{\scriptsize S}}}
\renewcommand{\ss}{{\mbox{\scriptsize s}}}
\newcommand{\R}{{\mbox{\scriptsize R}}}
\newcommand{\rr}{{\mbox{\scriptsize r}}}
\newcommand{\ret}{{\mbox{\scriptsize ret}}}
\newcommand{\schw}{{\mbox{\scriptsize schw}}}
\newcommand{\ext}{{\mbox{\scriptsize ext}}}
\newcommand{\ab}{_{ab}}
\renewcommand{\mu}{m}

\begin{document}
  \title{Gravitational radiation reaction and second-order
               perturbation theory}
  \author{Steven Detweiler }
  \affiliation{ Department of Physics,
            University of Florida,
            Gainesville, FL 32611-8440}
  \email{det@ufl.edu}
  \date{\today}

\begin{abstract} 
A point particle of small mass $\mu$ moves in free fall through a background
vacuum spacetime metric $g^0_{ab}$ and creates a first-order metric
perturbation $h^{1\ret}\ab$ that diverges at the particle. Elementary
expressions are known for the singular $\mu/r$ part of $h^{1\ret}\ab$ and for
its tidal distortion determined by the Riemann tensor in a neighborhood of
$\mu$. Subtracting this singular part $h^{1\s}\ab$ from $h^{1\ret}\ab$ leaves a
regular remainder $h^{1\R}\ab$. The self-force on the particle from its own
gravitational field adjusts the world line at $\Or(\mu)$ to be a geodesic of
$g^0_{ab}+h^{1\R}\ab$. The generalization of this description to second-order
perturbations is developed and results in a wave equation governing the
second-order $h^{2\ret}\ab$ with a source that has an $\Or(\mu^2)$ contribution
from the stress-energy tensor of $\mu$ added to a term quadratic in
$h^{1\ret}\ab$. Second-order self-force analysis is similar to that at first
order: The second-order singular field $h^{2\s}\ab$ subtracted from
$h^{2\ret}\ab$ yields the regular remainder $h^{2\R}\ab$, and the second-order
self-force is then revealed as geodesic motion of $\mu$ in the metric
$g^0\ab+h^{1\R}+h^{2\R}$.

\end{abstract}
  \maketitle

\section{Overview}
Recent, impressive fully relativistic numerical analysis has been brought to
bear on a black hole binary system with a mass ratio of 100 to 1
\cite{Lousto2011a,Lousto2011b}, and the evolution is followed for two full
orbits before coalescence. The two disparate length scales of an extreme or
intermediate mass-ratio binary pose a challenge for numerical relativists to
resolve the geometry in the vicinity of the small object while efficiently
analyzing the remainder of spacetime and providing gravitational wave trains
for a number of orbits. Second-order perturbation theory in general relativity
might more efficiently meet the challenge of the difficult numerical problems
of extreme and intermediate mass-ratio binaries.

Early descriptions of second-order perturbation theory \cite{Tomita74a,
Tomita74b, Campanelli99, Price00, Brizuela:2006ne, Brizuela:2007zza,
Brizuela:2009qd, Pazos:2010xf} have focused on perturbations with no matter
sources and are typically limited to metrics with a substantial amount of
symmetry.
 However, Habisohn \cite{Habisohn} presents a fully general description of
matter-free second-order perturbation theory for a background vacuum spacetime
metric $\gO\ab$.

Rosenthal \cite{Rosenthal05, Rosenthal06a, Rosenthal06b} was first to describe
a formal approach to second order perturbation theory which includes a
small-mass $\delta$-function point source. However, an actual application of
his approach does not appear to be straightforward.

The heart of this manuscript extends Habisohn's \cite{Habisohn} second-order
analysis to allow for a perturbing $\delta$-function point mass. Our formalism
is closely related to the traditional description of linear perturbation
theory.

We begin in Section \ref{habisohn} with the formal expansion of the Einstein
tensor, for a metric $g\ab+h\ab$, in powers of $h\ab$.
   First-order perturbation theory is summarized in Section \ref{1storder}
for the case that the source is a $\delta$-function object of small mass $\mu$.
In the test mass limit $\mu$ moves along a geodesic $\gm0$ of the background
metric $\gO\ab$. With a finite mass $\mu$ the metric is perturbed by the
retarded field $h^{1\ret}\ab$ at first order in $\mu$, and $\mu$'s worldline
deviates from $\gm0$ by an amount of $\Or(\mu)$ as $\mu$ itself interacts with
$h^{1\ret}\ab$ as a consequence of the first-order \textit{gravitational
self-force} as described in Section \ref{1stSF}.
 Throughout this manuscript we assume that the effects of $\mu$'s spin and
multipole structure on its motion are insignificant when compared with the
self-force effects.

The extension of Habisohn's \cite{Habisohn} second-order analysis to allow a
$\delta$-function point source demands careful consideration of the singular
behavior of the metric in a neighborhood of $\mu$ as described in Section
\ref{2ndorder}. Ultimately the wave equation for the second-order
$h^{2\ret}\ab$ appears in \Deqn{final} as one might have expected, and the
self-force analysis at second-order is seen to be similar in style to the
analysis at first-order.

The application of second order perturbation theory for a small mass still
requires an effort which is strongly dependent upon the details of the actual
problem of interest. Practical considerations are emphasized in Section
\ref{practical}.

\subsection*{Notation and conventions}

In a neighborhood of a geodesic $\gm0$ of the background metric $\gO\ab$ we use
\textit{locally inertial and Cartesian} (LIC) coordinates \cite{Zhang86} where
the timelike coordinate is $t$, the spatial indices $i$, $j$, $k$ and $l$ run
from 1 to 3, the spatial coordinates are $x^i$ and $r^2 \equiv x^i x^j
\eta_{ij}$. In addition LIC coordinates have special properties on $\gm0$: the
coordinate $t$ is the proper time, the spatial coordinates are all zero $x^i =
0$, the metric is the flat Minkowski metric $\eta\ab$, and all first coordinate
derivatives of $\gO\ab$ vanish. Second derivatives of $\gO\ab$ on $\gm0$
determine a curvature length and time scale $\calR$, and the components of the
Riemann tensor then scale as $1/\calR^2$ and their time derivatives along
$\gm0$ scale as $1/\calR^3$. After some fine-tuning of the coordinates
\cite{Zhang86,ThorneHartle, Det01}, the metric in a neighborhood of $\gm0$ may
be put into the form
\begin{align}
  \gO\ab\, dx^a\, dx^b &= \eta\ab\, dx^a\,dx^b
         -  x^i x^j R^0_{titj}( dt^2 + \delta_{kl}\, dx^k \, dx^l )
\nonumber\\ &\quad
   {} - \frac{4}{3}x^i x^j  R^0_{ikjt}\,dt\, dx^k + \Or(r^3/\calR^3) \, ,
\label{gOab}
\end{align}
where the superscript $^0$ on the components of the Riemann tensor implies that
it is to be evaluated on $\gm0$. Also, both $R^0_{titj}$ and $R^0_{ikjt}$ are
symmetric and tracefree in the indices $i$ and $j$ as consequences of the
vacuum Einstein equations.

Much of our analysis takes place in the \textit{buffer zone}
\cite{ThorneHartle}, a region spatially-surrounding $\gm0$ where $\mu \ll r \ll
\calR$. In the buffer zone $r$ is small enough compared to the curvature length
scale, $r\ll\calR$, that the curvature of $\gO\ab$ is barely apparent, and we
have the luxury of being able to expand the actual metric $\gO\ab+h^\ret\ab$
away from flat spacetime in powers of two simultaneously small numbers, $\mu/r$
and $r/\calR$.

\section{Expansion of the Einstein tensor}
\label{habisohn} We consider a perturbation $h\ab$ of a given metric $g\ab$,
and expand the Einstein tensor of the sum $G_{ab}(g+h)$ in terms of increasing
powers of $h\ab$ so that formally
\begin{align}
  G_{ab}(g+h) = G\ab(g) + G^{(1)}\ab(g,h) + G^{(2)}\ab(g,h) + \ldots
\label{expandG}
\end{align}
where Habisohn \cite{Habisohn} describes an individual term in this expansion
by
\begin{align}
  G^{(n)}_{ab}(g, h) = \frac{1}{n!}\left[\frac{\rd^n}{\rd\lambda^n}
                             G_{ab}(g+\lambda h)\right]_{\lambda=0}.
\label{expandg}
\end{align}
This notation implies that the operator $G^{(n)}_{ab}(g,h)$ returns an
expression that scales as $(h_{ab})^n$. For $n=1$ and $g\ab$ being a vacuum
solution of the Einstein equation,
\begin{align}
   2G^{(1)}_{ab}(g, h) &= -\nabla^c\nabla_c h_{ab} - \nabla_a \nabla_b h^c{}_c
           + 2 \nabla_{(a}\nabla^c h_{b)c}
\nonumber\\ & \quad
           - 2{R_a}^c{}_b{}^d h_{cd}
           + g_{ab}(\nabla^c\nabla_c  h^d{}_d - \nabla^c\nabla^d h_{cd}),
\label{Eab}
\end{align}
where $\nabla_a$ is the derivative operator compatible with the metric $g\ab$.
Habisohn \cite{Habisohn} provides the following expression for
$G^{(2)}_{ab}(g,h)$ in his Eq.~(3.1),
\begin{align}
  G^{(2)}_{ab}&(g,h) = \frac12h^{cd}\nabla_a \nabla_b h_{cd}
       +\frac14\left(\nabla_a h^{cd}\right)\nabla_b h_{cd}
\nonumber\\ & \quad
       + \left(\nabla^{[c}h^{d]}{}_a\right)\nabla_c h_{db}
       -\frac14 C^d\left(2\nabla_{(a} h_{b)d} - \nabla_d h_{ab} \right)
\nonumber\\ & \quad
       -h^{cd} \left(\nabla_c \nabla_{(a}h_{b)d}
                - \frac12\nabla_c \nabla_d h_{ab}\right)
\nonumber\\ & \quad + \left\{
      \frac18C^dC_d
      -\frac14h^{cd}\nabla^e\nabla_e h_{cd}
      -\frac18\left(\nabla^eh^{cd}\right)\nabla_eh_{cd}
      \right.
\nonumber\\ & \quad \left.
      +\frac14 h^{cd}\nabla_c C_d
      + \frac14 \left(\nabla^dh^{ce}\right) \nabla_c h_{de}
             \right\} g_{ab}
\end{align}
where
\begin{align}
   C_d \equiv 2 \nabla^ch_{cd} - \nabla_dh_c{}^c .
\end{align}

\section{First-order perturbation theory for a point mass}
 \label{1storder}
We next consider the consequences of adding an object of small size and small
mass $\mu$, with $\mu\ll \calR$, to the vacuum spacetime whose metric is
$\gO\ab$.

With a global coordinate system $(T,X^i)$, the stress-energy tensor for $\mu$
moving on a geodesic $\gm0$ of $\gO\ab$ is
\begin{align}
   T\ab(\gm0)
        & = \mu \frac{u_a u_b}{\sqrt{-\gO}}\frac{d\tau}{dT}
           \delta^3(X^i-\gm0^i(T)),
\label{Tabdef}
\end{align}
where  $\gm0^i(T)$ gives the spatial position of the geodesic as a function of
$T$, and the four-velocity $u_a$, $\sqrt{-\gO}$, and proper time $\tau$ are all
functions of $T$ along the worldline.

The dominant effect of $T\ab(\gm0)$ on the spacetime metric results in the
retarded metric perturbation $h^{1\ret}\ab$ proportional to $\mu$ which solves
\begin{align}
   G\ab(\gO+h^{1\ret}) = 8 \pi T\ab(\gm0) + O(\mu^2) \, ,
\label{Gab8piTab}
\end{align}
with appropriate boundary conditions. The superscript 1 on any metric
perturbation implies that $h^{1\ret}_{ab}$ is $\Or(\mu)$, for example. Later we
use $h^{2\ret}\ab$ for an $\Or(\mu^2)$ metric perturbation and also use
$h^{\ret}\ab \equiv h^{1\ret}\ab + h^{2\ret}\ab +O(\mu^3)$.

For this linear perturbation problem, we expand the Einstein tensor in
\Deqn{Gab8piTab} using \Deqn{expandG} and isolate the terms linear in $\mu$  to
obtain the first-order perturbation equation,
\begin{align}
  G^{(1)}\ab(\gO, h^{1\ret}) = 8 \pi T\ab(\gm0) .
\label{1stOrd}
\end{align}

The Bianchi identity implies for arbitrary $h\ab$ that if $g_{ab}$ is a vacuum
solution of the Einstein equation, then
\begin{align}
    \nabla^aG^{(1)}_{ab}(g, h)= 0,
\label{linearbianchi}
\end{align}
perhaps as a distribution.  An integrability condition for \Deqn{1stOrd} thus
requires that $T\ab(\gm0)$ be divergence free. The assumption that the
worldline of $\mu$ is a geodesic $\gm0$ of $\gO\ab$ guarantees that
 $\nabla^a T\ab(\gm0) = 0$ and
that the integrability condition is satisfied.

\section{First-order gravitational self-force}
 \label{1stSF}
After $h^{1\ret}\ab$ is found using \Deqn{1stOrd} there are several ways of
calculating, understanding and interpreting the gravitational self-force \cite{
Det01, Mino97, DetWhiting03, Det05, PoissonPVLR, QuinnWald97, Gralla08}.
 Our favorite is to note that $h^{1\ret}\ab$ is naturally
decomposed within a neighborhood of $\gm0$ into two complementary parts,
\begin{align}
  h^{1\ret}\ab = h^{1\s}\ab + h^{1\R}\ab.
\label{decomp1}
\end{align}

The first part $h^{1\s}\ab$ is the linear piece of the
\textit{singular field} 
$h^{\s}\ab$ which is a special solution of
\begin{align}
   G\ab(\gO + h^{\s}) = 8\pi T\ab(\gm0)
\label{singdef}
\end{align}
with the notable features that $h^\s\ab$:
  (1) may be expanded in powers of $\mu$,
  (2) is local to $\mu$ and does not depend upon boundary conditions,
  (3) is accessible via an asymptotic expansion
  \cite{Det01,Mino97,DetWhiting03,Det05,PoissonPVLR}
  each term of which is singular or of limited differentiability on $\gm0$,
  and (4) does not exert a force on $\mu$ itself, just as the Coulomb field of
  an electron at rest exerts no net force on the electron.

The substitution $h^{\s}\ab = h^{1\s}\ab + h^{2\s}\ab+O(\mu^3)$, with
$h^{2\s}\ab =
\Or(\mu^2)$, into \Deqn{singdef} 
and the expansion of the Einstein tensor results in two equations, the first
linear in $\mu$ and the second quadratic,
\begin{align}
   G^{(1)}\ab(\gO, h^{1\s}) &= 8\pi T\ab(\gm0)
\label{firstorderS}
\\
   G^{(1)}\ab(\gO, h^{2\s}) &= - G^{(2)}\ab(\gO, h^{1\s}) .
\label{secondorderS}
\end{align}

The inhomogeneous, linear singular field $h^{1\s}\ab$ looks like a Coulomb
$\mu/r$ field being tidally distorted by the Riemann tensor of $\gO\ab$.  We
qualitatively describe  $h^{1\s}\ab$, using LIC coordinates associated with
$\gm0$, as
\begin{align}
  h^{1\s}\ab\sim \frac{\mu}{r} \left( 1+ \frac{x^2}{\calR^2} + \ldots\right)
      \, ;
\label{h1Sbare}
\end{align}
only the scaling of the leading terms are shown, and this scaling is valid in
the buffer zone, where $\mu \ll r \ll \calR$. We distinguish $x$ from $r$ to
emphasize that $x/r$ is generally finite but discontinuous $C^{-1}$ in the
limit $r\rightarrow0$. The dominant term, scaling as just $\mu/r$, represents
the linear in $\mu$ terms in an $\mu/r$ expansion of the Schwarzschild metric,
as given in \Deqn{h1abS} in Appendix \ref{smallhole}. The second term in the
parentheses reflects the quadrupole distortion of the $\mu/r$ field that is
induced by the external Riemann tensor's tidal effects which scale as
$x^2/\calR^2$, as given by the terms proportional to $m$ in \Deqn{h2ab}.

The complement of $h^{1\s}\ab$ is the homogeneous \textit{regular} field
$h^{1\R}\ab = h^{1\ret}\ab- h^{1\s}\ab$, from \Deqn{decomp1}, which solves
\begin{align}
   G^{(1)}\ab(\gO, h^{1\R}) = 0 \, .
\label{h1Rhomog}
\end{align}
The regular field $h^{1\R}\ab$ is smooth on $\gm0$ and, thus, qualitatively
described in a neighborhood of $\gm0$ by
\begin{align}
  h^{1\R}\ab\sim \frac{\mu}{\calR} + \frac{\mu x}{\calR^2}
           + \frac{\mu x^2}{\calR^3} + \ldots
      \; ,
\label{h1Rbare}
\end{align}
with the LIC coordinates associated with $\gm0$. Each term takes the form of an
external multipole moment proportional to $\mu$.

The regular field $h^{1\R}\ab$ is added to 
$\gO\ab$ to create the \textit{external} metric
\begin{equation}
  g^\ext\ab \equiv \gO\ab + h^{1\R}\ab
\label{externalMetric}
\end{equation}
which governs the geodesic motion of $\mu$.
  After all, $h^{1\R}\ab$ is a homogeneous solution of
\Deqn{h1Rhomog} with no variation over a length scale comparable to $\mu$. An
observer in a neighborhood of $\mu$, with no a priori knowledge of the global
spacetime, could measure the actual metric
  $\gO\ab + h^{1\R}\ab + h^{1\s}\ab$ at $\Or(\mu)$ and could distinguish the
singular behavior of $h^{1\s}\ab$ from the remainder $\gO\ab+h^{1\R}\ab$.
However, the observer would be unable to distinguish $h^{1\R}\ab$ from $\gO\ab$
in the combination $\gO\ab+h^{1\R}\ab$ at linear order via \textit{local
measurements only} because $\gO\ab+h^{1\R}\ab$ is a smooth solution of the
vacuum Einstein equations at linear order. The observer would then naturally
note that the worldline of $\mu$ is a geodesic $\gm0+\gm{1\R}$ of the metric
$\gO\ab+h^{1\R}\ab$. The difference between the two worldlines
is denoted $\gm{1\R}$ and reflects the effects of what is often called the
gravitational self-force, even though there is neither a force on $\mu$ nor an
acceleration of its worldline within the external metric $\gO\ab+h^{1\R}\ab$.

It is apparent that an $\Or(\mu)$ coordinate transformation of the original LIC
coordinates for $\gm0$ would remove the dipole term in \Deqn{h1Rbare} and put
the sum $\gO\ab + h^{1\R}\ab$ into the same form as displayed in \Deqn{gOab},
with $\Or(\mu)$ changes in the components of the \textit{external} Riemann
tensor. 

In an application $h^{1\ret}\ab$ is typically found numerically while
$h^{1\s}\ab$ (or its approximation, {\it cf.} Section \ref{practical}) is found
analytically, then $h^{1\R}\ab = h^{1\ret}\ab - h^{1\s}\ab$ gives the regular
remainder (or its approximation) which is used to determine the self-force and
the appropriate geodesic $\gm0+\gm{1\R}$ of $\gO\ab+h^{1\R}\ab$.

\section{Second-order perturbation theory}
 \label{2ndorder}
We assume that we have solved a first-order self-force problem of interest and
have, in hand,  $h^{1\ret}_{ab}$, $h^{1\s}_{ab}$, $h^{1\R}_{ab}$, the initial
geodesic $\gm0$ of $\gO_{ab}$ and the self-force modified geodesic
$\gm0+\gm{1\R}$ of $\gO_{ab}+h^{1\R}_{ab}$.

For the second order problem we also require $h^{2\s}\ab$ which can be
determined via an asymptotic expansion of \Deqn{secondorderS}, and scales as
\begin{align}
  h^{2\s}\ab\sim \frac{\mu^2}{r^2} \left( 1+ \frac{x^2}{\calR^2}
    + \ldots \right)
\label{h2Sbare}
\end{align}
 with LIC coordinates.
  The dominant term, scaling as $\mu^2/r^2$, is the term quadratic in $\mu$ in an
$\mu/r$ expansion of the Schwarzschild metric, as given in \Deqn{hnabS} for
$n=2$.
   The second term in the parentheses reflects the quadrupole distortion of the
$\mu^2/r^2$ field that is induced by the external Riemann tensor's tidal
effects which scale as $x^2/\calR^2$, as given by the $O(m^2)$ terms in
\Deqn{h2ab}.

To understand second-order perturbation theory requires understanding two
distinct and critical roles played by the first order regular field
$h^{1\R}\ab$.
 First, the stress-energy tensor of $\mu$ is $T\ab(\gm0+\gm{1\R})$, where the
argument implies that the worldline of $\mu$ is now a geodesic of $\gO_{ab} +
h^{1\R}_{ab}$. The change in the stress-energy tensor resulting from the
first-order self-force is
\begin{align}
   T\ab(\gm0+&\gm{1\R}) - T\ab(\gm0)
 \nonumber\\ &
      = \mu\Delta\left(\frac{u_a u_b}{\sqrt{-g}}\frac{d\tau}{dT}\right)
           \delta^3[X^i-\gm0^i(T)]
 \nonumber\\ & \quad
  - \mu\frac{u_a u_b}{\sqrt{-\gO}}\frac{d\tau}{dT}
         \;\gm{1R}^j\frac{\partial}{\partial X^j} \delta^3[X^i-\gm0^i(T)]
\label{DeltaTab}
\end{align}
where the $\Delta$ operation reflects the $\Or(\mu)$ change in the quantity in
parentheses which follows from changing the metric to $\gO\ab+h^{1\R}\ab$ from
$\gO\ab$. Thus the difference between the two stress-energy tensors is a
distribution of $\Or(\mu^2)$ with support on $\gm0$ and consists of terms with
a $\delta$-function and with a gradient of a $\delta$-function.

A second effect of $h^{1\R}\ab$ on the second-order problem is the modification
of the tidal environment of $\mu$ by $h^{1\R}_{ab}$ which becomes an $\Or(\mu)$
part of the \textit{external} metric as in \Deqn{externalMetric}. This creates
$\Or(\mu)$ changes in the the external Riemann tensor's multipole moments.
These changes are responsible for $O(\mu^2)$ corrections to $h^{1\s}\ab$ which
we label $h^{2\s\dag}\ab$.
 Thus the singular field is \textit{not} derived solely from the initial
geodesic and the background metric $\gO_{ab}$, rather it specifically includes
effects from the self-force modification of the geodesic and from the
additional $\Or(\mu^2)$ tidal distortion of $h^{1\s}\ab$ caused by
$h^{1\R}_{ab}$, and these $O(\mu^2)$ contributions to the singular field
constitute $h^{2\s\dag}\ab$.

The presence of $h^{1\R}\ab$ in the external metric $\gO\ab+h^{1\R}\ab$
modifies the tidal effects of the external Riemann tensor on the singular field
and \Deqn{h1Sbare} becomes
\begin{align}
  h^{1\s}\ab + h^{2\s\dag}\ab\sim \frac{\mu}{r} \left[ 1 +
    \frac{x^2}{\calR^2}\left(1+\frac{\mu}{\calR}\right)
    + \ldots \right],
\label{h1Smod}
\end{align}
where we are now using LIC coordinates for the geodesic $\gm0+\gm{1\R}$ of
$\gO\ab+h^{1\R}\ab$. The $\mu/\calR$ term in the parentheses adds an
$\Or(\mu^2)$ contribution to $h^{\s}\ab$; however, the $O(\mu^2)$ $h^{2\s\dag}$
is naturally grouped with $h^{1\s}\ab$ because its presence in \Deqn{h1Smod}
algebraically resembles part of $h^{1\s}\ab$ in \Deqn{h1Sbare} much more than
any part of $h^{2\s}\ab$ in \Deqn{h2Sbare}.

Through second order the singular field is thus represented by
\begin{align}
  h^\s_{ab} &= h^{1\s}_{ab} + h^{2\s\dag}_{ab} + h^{2\s}_{ab} + O(\mu^3).
\label{hS2nd}
\end{align}

An immediate application of this notation is in the recognition that
\begin{align}
 G\ab^{(1)}(\gO+h^{1\R}, h^{1\s} + h^{2\s\dag})
       = 8\pi T\ab(\gm0 + \gm{1\R}) + \Or(\mu^3),
\label{1stOrdstar}
\end{align}
which is the natural extension of \Deqn{firstorderS} to second-order.
 The presence of $h^{1\R}\ab$ as part of the external metric in the first
argument of $G\ab^{(1)}$ requires the addition of $h^{2\s\dag}\ab$ to the
second argument. We have already described $h^{1\R}\ab$ in \Deqn{decomp1}, and
it is natural then to define $h^{2\R}\ab$ via
\begin{align}
   h^{2\ret}\ab &= h^{2\R}\ab + h^{2\s\dag}\ab + h^{2\s}\ab .
\end{align}

We now confront the second-order problem which requires a solution for
$h^{2\ret}\ab$ from
\begin{align}
  G\ab(\gO + h^{1\ret} + h^{2\ret})
       = 8\pi T\ab(\gm0+\gm{1\R}) + \Or(\mu^3) ,
\label{2ndOrdA}
\end{align}
when we are given the metric perturbations $h^{1\ret}\ab$, $h^{1\R}\ab$,
$h^{1\s}\ab$, $h^{2\s\dag}\ab$, $h^{2\s}\ab$, and the worldlines $\gm0$ and
$\gm0+\gm{1\R}$.
  We expand the left hand side about $\gO\ab$, rearrange some terms, and
substitute for $G^{(1)}\ab(\gO,h^{1\ret})$ from \Deqn{1stOrd} to obtain
\begin{align}
  G^{(1)}\ab(\gO,h^{2\ret})
       & = 8\pi T\ab(\gm0+\gm{1\R}) - 8\pi T\ab(\gm0)
\nonumber\\ &\quad - G^{(2)}\ab(\gO,h^{1\ret}) \; .
\label{final}
\end{align}
This wave equation for $h^{2\ret}\ab$ is the primary formal result of this
manuscript.
 At the source each stress-energy term is $\Or(\mu)$; however, their difference
is a distribution with support on $\gm0$ and is of $\Or(\mu^2)$ as given in
\Deqn{DeltaTab}.

The integrability condition for \Deqn{final} is easily satisfied away from
$\gm0$ because there $G^{(1)}\ab(\gO,h^{1\ret})=0$ and the fact that for any
$h\ab$ if $G\ab^{(1)}(\gO, h)= 0$ then it follows that $\nabla^a
G^{(2)}_{ab}(\gO, h)= 0$, as shown by Habisohn \cite{Habisohn} in his
Eq.~(3.7). Thus the divergence of the right hand side is zero away from $\gm0$.
The discussion of the integrability condition in a neighborhood of $\gm0$ is
deferred until just after \Deqn{h2reqn} below.

\begin{widetext}
\Deqn{final} becomes surprisingly transparent after some analysis (while
cavalierly dropping terms of $O(\mu^3)$ along the way) when $h^{\ret}\ab$ is
re-expressed with the substitutions $h^{1\ret}\ab = h^{1\R}\ab + h^{1\s}\ab$
and $h^{2\ret}\ab = h^{2\R}\ab + h^{2\s}\ab + h^{2\s\dag}\ab$.
  Then the substitutions for the stress-energy tensors from \Deqns{firstorderS}
and (\ref{1stOrdstar}) lead to
\begin{align}
G^{(1)}\ab(\gO,h^{2R}&+h^{2\s\dag}+h^{2\s})  =
G^{(1)}\ab(\gO+h^{1R}, h^{1\s}+h^{2\s\dag}) - G^{(1)}\ab(\gO,h^{1\s})
         - G^{(2)}\ab(\gO,h^{1R}+h^{1\s}) \; .
\end{align}
Use of the identity in \Deqn{BigIdentity} modifies the RHS with the result that
\begin{align}
 G^{(1)}\ab(\gO,h^{2R} & +h^{2\s\dag}+h^{2\s})  =
   G^{(1)}\ab(\gO+h^{1R}, h^{1\s}+h^{2\s\dag})
  - G^{(2)}\ab(\gO,h^{1\s})
 - G^{(2)}\ab(\gO,h^{1R}) - G^{(1)}\ab(\gO+h^{1R},h^{1\s}) .
\end{align}
\end{widetext}
On the RHS, the fourth term cancels that part of the first term which is linear
in $h^{1\s}\ab$.
  The terms linear in $h^{2\s}$ on the LHS and quadratic in $h^{1\s}$ on the RHS
cancel from \Deqn{secondorderS}.
  The terms linear in $h^{2\s\dag}$ on each side of the equation cancel up to a
term of $O(\mu^3)$, which is ignored.
  When the dust has settled what remains is
\begin{align}
  G^{(1)}\ab(\gO,h^{2\R}) = - G^{(2)}\ab(\gO,h^{1\R}),
\label{h2Reqn}
\end{align}
which reveals obvious consistency for this second order perturbation formalism:
When $h^{1\s}\ab$, $h^{2\s\dag}\ab$ and $h^{2\s}\ab$ correctly capture their
respective parts of the singular behavior of the retarded field, the regular
remainder $h^{1\R}\ab+h^{2\R}\ab$ appears as a source-free metric perturbation
at second order in $\mu$ as described by Habisohn \cite{Habisohn}.
 The integrability condition for \Deqn{h2Reqn} is satisfied in a manner similar to
that for \Deqn{final} away from $\gm0$.

The second-order self-force is similar to the first-order self-force. In a
neighborhood of $\mu$, $h^{2\ret}\ab$ is naturally decomposed into two
complementary parts, $h^{2\ret}\ab = h^{2\R}\ab + (h^{2\s\dag}\ab+h^{2\s}\ab)$,
where  $h^{2\s\dag}\ab+h^{2\s}\ab$ exerts no force on $\mu$ itself. The
second-order self-force then moves $\mu$ along a geodesic of $\gO\ab +
h^{1\R}\ab + h^{2\R}\ab$.

The sanguine simplicity of \Deqn{h2Reqn} hides the complexity of its
application. It might appear as though $h^{2\R}\ab$ may be solved only in terms
of $h^{1\R}\ab$ in a neighborhood of $\gm0$, but what is lacking is the
description of the boundary condition which is typically given as a condition
on the retarded field $h^\ret\ab$. To find $h^{2\R}\ab$ it is necessary first
to find $h^{1\ret}\ab$ and to evaluate $h^{1\s}\ab$ as an asymptotic expansion
in a neighborhood of $\gm0$; these lead to $h^{1\R}\ab = h^{1\ret}\ab -
h^{1\s}\ab$. With $h^{1\R}\ab$ the self-force modification of the worldline may
be determined. At this point $h^{2\s\dag}\ab$ and $h^{2\s}\ab$ are accessible
via asymptotic expansions and $h^{2\ret}\ab$ could be evaluated via
\Deqn{final}. Only then is $h^{2\R}\ab$ able to be determined.

\section{Practical concerns}
 \label{practical}
In most situations, only an asymptotic approximation $h^\ss\ab$ to the exact
$h^\s\ab$ is likely to be known, and as a consequence an actual application of
the formalism described above is not as elementary as it might appear. In this
case, $h^r\ab \equiv h^\ret\ab - h^\ss\ab$ is an approximation to the actual
regular field $h^R\ab$.
 With these approximations some concerns appear in a neighborhood of the
$\delta$-function point source $\mu$. The proper evaluation of the self-force,
via $h^r\ab$, requires that $h^r\ab$ match both the value and first coordinate
derivatives of $h^\R\ab$ on $\gm0$.
 In turn, this requires that the difference $h^\s\ab - h^\ss\ab$ be zero on
$\gm0$, and also, with LIC coordinates, that all first coordinate derivatives
of this difference also be zero on $\gm0$.

Experience \cite{DetMessWhiting03, RiveraMessaritakiWhitingDet04, VegaDet08,
Vega11, VegaDienerTD09, Diener:2011cc, Wardell:2011gb} has shown that in
numerical work if the difference $h^\s\ab - h^\ss\ab$ of these two singular
fields is increasingly more differentiable, then the numerical analysis will be
increasingly more accurate.

In some self-force analyses \cite{det08}
\begin{align}
    h^{1\s}\ab & = h^{1\ss}\ab+ O(\mu x^4/r\calR^4) \;\; \text{and}
\nonumber\\
    h^{2\s\dag}\ab+h^{2\s}\ab &= h^{2\ss\dag}\ab+h^{2\ss}\ab+\Or(\mu^2 x^4/r^2\calR^4).
\end{align}
We assume henceforth that we have such a precisely described approximation
$h^\ss\ab$ to $h^\s\ab$.

For first order analyses, the integrability condition required for using
\Deqn{1stOrd} to solve for $h^{1\ret}\ab$ is easily satisfied. The
approximation for $h^{1\ss}\ab$ is then accurate enough that $h^{1\rr}$ is
$C^2$ on $\gm0$, and the accuracy of the computed self-force effects are not
limited by this approximation.

To derive a second-order equation for $h^{2\rr}\ab$ follow the same
instructions as for \Deqn{h2Reqn} while using $h^\rr\ab$ and $h^\ss\ab$ instead
of $h^\R\ab$ and $h^\s\ab$, and do not use \Deqns{firstorderS},
(\ref{secondorderS}) or (\ref{1stOrdstar}) for substitutions. The result is
\begin{align}
  G^{(1)}\ab(\gO, h^{2\rr}) & = - G^{(2)}\ab(\gO,h^{1\rr})
\nonumber\\ &\quad
      - [G^{(2)}\ab(\gO,h^{1\ss}) + G^{(1)}\ab(\gO,h^{2\ss})]
\nonumber\\ &\quad
  \hskip-.5in + [8\pi T_{ab}(\gm0+\gm{1\rr})
                  - G^{(1)}\ab(\gO+h^{1\rr}, h^{1\ss}+ h^{2\ss\dag})]
\nonumber\\ &\quad
      - [8\pi T_{ab}(\gm0) - G^{(1)}\ab(\gO, h^{1\ss})] .
\label{h2reqn}
\end{align}

The integrability condition for using \Deqn{h2reqn} to solve for $h^{2\rr}$ is
satisfied everywhere except, perhaps, precisely on $\gm0$ where the analysis
entails some modest difficulty.
  The order terms associated with $h^{1\ss}\ab$ and $h^{2\ss\dag}\ab+h^{2\ss}\ab$
(given above) provide an estimate for the behavior of the source on the
righthand side in a neighborhood of $\gm0$. Most of the terms on the righthand
side are either distributions or differentiable and well behaved on $\gm0$. The
uncertainty involving the source is dominated by the $G^{(2)}\ab(\gO,h^{1\ss})$
and $G^{(1)}\ab(\gO,h^{2\ss\dag}+h^{2\ss})$ terms; each of
  these scales as two spatial derivatives of $\mu^2 x^4/r^2\calR^4$, which is
$\Or(\mu^2 x^2/r^2\calR^4)$ and finite but discontinuous on $\gm0$.
  The divergence of this term is then $\Or(\mu^2x/r^2\calR^4)$ which diverges on
$\gm0$. However, the integral of this divergence (contracted with a smooth test
vector field of order unity) over a small volume of radius $r_*$ about $\mu$ is
then $\Or(\mu^2r_*^2/\calR^4)$. If we choose $r_*$ such that $\mu$, $r_*$, and
$\calR$ are related by
\begin{equation}
 r_*^2/\calR \lesssim  \mu \ll r_* \ll \calR \; ,
\end{equation}
then it follows that the integrated divergence over the volume of radius $r_*$
is $\Or(\mu^3/\calR^3)$. For $r > r_*$ the integrability condition is
satisfied. Thus, the integrability condition fails only at $\Or(\mu^3)$ which
does not hinder the analysis at $\Or(\mu^2)$. No fundamental difficulty
prevents solving \Deqn{h2reqn} for $h^{2\rr}\ab$. The resultant $h^{2\rr}\ab$
is $C^1$ on $\gm0$ and is sufficient to find second order self-force effects.

\section{Summary and conclusions}

Upon reflection, \Deqn{final} describes the second-order perturbation problem
for a $\delta$-function point mass in a quite satisfactory manner and is the
primary result of this manuscript. The metric perturbation $h^{2\ret}\ab$ may
be determined directly, and the $h^\s\ab$, $h^\R\ab$ decomposition of
$h^\ret\ab$ is only required for determining the effects of the self-force.

It is notable that the representation of a small mass $\mu$ by a
$\delta$-function point source works as well at second-order as it does at
first order.

\section{Acknowledgment} I am grateful for insightful discussions with Ian
Vega and Bernard Whiting (particularly for his suggestion to pursue a wave
equation for $h^{2\ret}\ab$ rather than for $h^{2\rr}\ab$) and for the
encouraging atmosphere of the greater Capra Community. This work was supported
in part by the National Science Foundation under grant PHY-0855503 with the
University of Florida.

\appendix

\section{Nonlinear perturbation theory and tidal distortion of a small black hole}
 \label{smallhole}

The simplest example of non-linear perturbation theory in General Relativity
involves perturbing flat spacetime by putting a small, spherical object of mass
$\mu$ down on the origin of Minkowski space. Outside the object the geometry
must be the Schwarzschild metric from Birkhoff's theorem.

The usual coordinates of Minkowski space form an LIC coordinate system because
the spatial origin $x^i=0$ is a geodesic, and the other LIC conditions are
clearly satisfied. We define a covariant vector in the radial direction via
$n_i = \nabla_i r$. With a Schwarzschild black hole of mass $\mu$ present at
the spatial origin, the metric takes the unfamiliar form
\begin{align}
  g^\schw_{ab}\dx^a \dx^b &= -\left(1-\frac{2\mu}{r}\right)\dt^2
      + \frac{r}{r-2\mu}n_k n_l \dx^k \dx^l
\nonumber\\ & \quad  + (\delta_{kl}- n_k n_l) \dx^k dx^l \, .
\label{Schw}
\end{align}
 An alternative description of this form of the Schwarzschild metric is
\begin{align}
   g^\schw_{ab} = \eta_{ab} + {}_0h^\s_{ab} \, ,
\label{0hdef}
\end{align}
where ${}_0h^\s_{ab}$ is to be identified as the \textit{singular field} from
self-force analysis, and the leading subscript 0 implies that this monopole
part of the singular field is spherically symmetric.
 From \Deqn{Schw} it follows that
\begin{align}
  {}_0h^\s\ab \dx^a \dx^b &= (g^\schw_{ab}-\eta_{ab})\dx^a \dx^b
\nonumber\\
     & = \frac{2\mu}{r}\dt^2
      + \frac{2\mu}{r-2\mu}n_k n_l \dx^k \dx^l \, .
\end{align}
The $n$th order part of ${}_0h^\s_{ab}$ scales as $\mu^n$ and may be isolated
with
\begin{align}
   {}_0h^{n\s}_{ab} \equiv \frac{\mu^n}{n!}
       \left[\frac{\rd^n}{\rd\mu^n} h^\s_{ab} \right]_{\mu=0} \; .
\label{hmuscale}
\end{align}
This provides the formal representation
\begin{align}
  h^\s_{ab} = \sum_{n=1}^\infty h^{n\s}_{ab} \;.
\end{align}
For our elementary example, the first term in this sum is
\begin{align}
  {}_0h^{1\s}\ab \dx^a \dx^b & =
     \frac{2\mu}{r}\dt^2
      + \frac{2\mu}{r}n_k n_l \dx^k \dx^l ,
\label{h1abS}
\end{align}
and for $n>1$
\begin{align}
  {}_0h^{n\s}\ab \dx^a \dx^b & =
            \left(\frac{2\mu}{r}\right)^n n_k n_l \dx^k \dx^l .
\label{hnabS}
\end{align}
In this treatment of the Schwarzschild metric the singular features of
${}_0h^{n\s}\ab$ are identified, and the absence of a regular field $h^\R\ab$
is assured by the flat nature of the initial Minkowski metric.

A more subtle example places a Schwarzschild black hole in a region of
spacetime that is empty but has slowly changing curvature from some distant
source. In that case the metric of a black hole placed on the origin of the LIC
coordinate system of \Deqn{gOab} would be perturbed by the background curvature
and could be analyzed by use of the Regge-Wheeler \cite{ReggeWheeler}
formalism.
   The boundary condition at large $r$ requires that the perturbed metric
approach the form given in \Deqn{gOab}. The boundary condition as
$r\rightarrow2\mu$ requires that the perturbation be well behaved on the future
event horizon of the small black hole.
 In the time independent limit the wave equations for the metric perturbations
admit analytic solutions which satisfy the boundary conditions \cite{Det01}.

The dominant tidal effects present in both $h^{1\s}\ab$ and $h^{2\s}\ab$ are
seen in the quadrupole $l=2$ terms of Eq.~(9) of \cite{Det01}, which we
reproduce here as
\begin{eqnarray}
  {}_2h^\s_{ab}dx^a dx^b & = &
         R^0_{titj}x^i x^j \big[ (4\mu/r-4\mu^2/r^2)\, \dt^2
\nonumber  \\ & &
         {} + 2\mu^2/r^2 (\delta_{kl}-n_k n_l)\,\dx^k \dx^l \big]
\nonumber  \\ & &
         {} + \frac{8\mu}{3r} x^i x^j R^0_{ikjt} \,\dx^k \dt
         + \Or(m x^3/r\calR^3)
\nonumber  \\ & &
         + \Or(m^2 x^2/r\calR^3) + \Or(m^3 x^2/r^2 \calR^3)
\label{h2ab}
\end{eqnarray}
The order terms here result from the possible slow time dependence of the tidal
field and are all much smaller in the buffer zone than the explicit terms
provided.

A more extensive analysis of $h^\s\ab$ in a similar style is given in
\cite{Det05}. An alternative treatment in a dramatically different style is
given in \cite{PoissonPVLR}.

\section{Useful Identity}
An identity used in deriving \Deqns{h2Reqn} and (\ref{h2reqn}) results from
considering two different expansions of the same expression $G(\gO + h^{1\R} +
h^{1\s})$. On the one hand, treating $h^{1\R}\ab + h^{1\s}\ab$ as a single
quantity, it expands to be
\begin{align}
  G\ab^{(1)}(\gO,h^{1\R} + h^{1\s})
     + G\ab^{(2)}(\gO,h^{1\R} + h^{1\s}) + \Or(\mu^3) .
\label{appG2eqnA}
\end{align}
On the other hand, first grouping $h^{1\R}\ab$ with $\gO\ab$ while expanding in
powers of $h^{1\s}\ab$, and subsequently expanding in powers of $h^{1\R}\ab$,
it becomes
\begin{align}
  G\ab^{(1)}(\gO,h^{1\R}) & + G\ab^{(2)}(\gO,h^{1\R})
     + G\ab^{(2)}(\gO, h^{1\s})
\nonumber\\ &
    + G\ab^{(1)}(\gO+h^{1\R}, h^{1\s}) + \Or(\mu^3) .
\label{appG2eqnB}
\end{align}
Equating these two expressions reveals that
\begin{align}
   G\ab^{(2)}(\gO,h^{1\ret}) &=
      G\ab^{(2)}(\gO, h^{1\s}) + G\ab^{(2)}(\gO,h^{1\R})
\nonumber\\
   &\quad {} + G\ab^{(1)}(\gO+h^{1\R}, h^{1\s}) -  G\ab^{(1)}(\gO, h^{1\s})
\nonumber\\
   &\quad {} + \Or(\mu^3) .
\label{BigIdentity}
\end{align}

\bibliographystyle{prsty}


\end{document}